\documentclass[%
superscriptaddress,
 amsmath,amssymb,
 aps,
 twocolumn,
 prl,
 longbibliography,
]{revtex4-2}

\usepackage{xcolor}
\usepackage{float}
\usepackage{braket}

\usepackage{graphicx}
\usepackage{dcolumn}
\usepackage{bm}

\begin{document}

\title{Sound Waveguiding by Spinning: An Avenue towards Unidirectional Acoustic Spinning Fibers}

\author{Mohamed Farhat}
\affiliation{Computer, Electrical, and Mathematical Science and Engineering (CEMSE) Division, King Abdullah University of Science and Technology (KAUST), Thuwal 23955-6900, Saudi Arabia}


\author{Pai-Yen Chen}
\affiliation{Department of Electrical and Computer Engineering, University of Illinois at Chicago, Chicago, Illinois 60607, USA}

\author{Ying Wu}
\email{ying.wu@kaust.edu.sa}
\affiliation{Computer, Electrical, and Mathematical Science and Engineering (CEMSE) Division, King Abdullah University of Science and Technology (KAUST), Thuwal 23955-6900, Saudi Arabia}

\date{\today}

\begin{abstract}
Waveguiding in general and acoustic waveguiding in particular are possible at the condition of having a transverse "discontinuity" or modulation of the refractive index. We propose here a radically different approach that relies on imposing spinning on a column of air, leading to high modified acoustic refractive indices for specific azimuthal modes. Such discovery may be leveraged to realize not only the airborne acoustic counterpart of the optical fiber, i.e., the acoustic spinning fiber (ASF), but also nonreciprocal unidirectional waveguiding mechanism, reminiscent of the ''acoustic Zeeman effect''. The concept is demonstrated in the realm of acoustics, yet it can be applicable to other wave systems, e.g., photonics or elastodynamics.
\end{abstract}

\maketitle


Metasurfaces and metamaterials consisting of periodically distributed unit-cells (or meta-atoms) in two or three dimensions (2D, 3D), respectively \cite{cai2010optical}, have revolutionized the way to control optical wave propagation that led to broad applications ranging from superlensing \cite{fang2003regenerating}, invisibility cloaks \cite{chen2010transformation}, optical waveguides \cite{kadic2012transformation}, to computing \cite{silva2014performing}. Such concepts were further extended to other fields, including but not limited to, acoustic \cite{assouar2018acoustic,chen2011acoustic,dupont2011numerical}, elastic \cite{mei2003theory,farhat2014platonic}, structural \cite{lapine2009structural}, and thermodynamical \cite{qu2017micro} waves.
In particular, acoustic waveguiding for airborne sound requires a relative acoustic refractive index greater than one, or equivalently the speed of sound lower than that in air at room temperature (343 m/s) \cite{morse1968ku}, which is almost impossible to be realized with normal materials. Metamaterial structures may result in an effective refractive index with the required behavior as reported in Ref. \cite{zangeneh2018acoustic}, for example. But such route is complicated to realize (subwavelength structuring) and the effect is limited to a narrow-band operation. Recently, in optics, a different strategy was suggested that exploits the vectorial spin-orbit interaction of light and its geometric phases \cite{slussarenko2016guiding}. Yet, this approach is also difficult to implement in acoustics, as it requires anisotropic media with tailored optical axes. The practical realization of airborne sound waveguiding remains a grand challenge.

Recently, time-varying metamaterials were proposed as a promising paradigm \cite{caloz2019spacetime,caloz2019spacetime2} for designing intriguing functionalities such as luminal devices \cite{huidobro2019fresnel} and/or amplifying structures \cite{galiffi2019broadband,farhat2021spacetime}. In the same vein, acoustic metamaterials with moving components have been brought up to enrich the application spectrum \cite{censor1971scattering,censor1973two,graham1969effect,schoenberg1973elastic,ramaccia2017doppler,zhao2020acoustic}. For instance, spinning acoustic fluids were shown to possess scattering cancellation properties \cite{farhat2020scattering} and to lead to negative acoustic torque and force without the transverse variation of the refractive index \cite{farhat2021transverse}. These metamaterials with spinning components were also shown to break reciprocity \cite{fleury2014sound,quan2021odd} or to lead to topological acoustics \cite{yang2015topological}. These achievements naturally inspired us to consider the spinning of air columns as a candidate to realize the confinement and waveguiding of sound, which obeys a modified (azimuthal-dependent) Helmholtz-like equation with a modified wavenumber \cite{farhat2020scattering}. 

\begin{figure}[b!]
    \centering
    \includegraphics[width=\columnwidth]{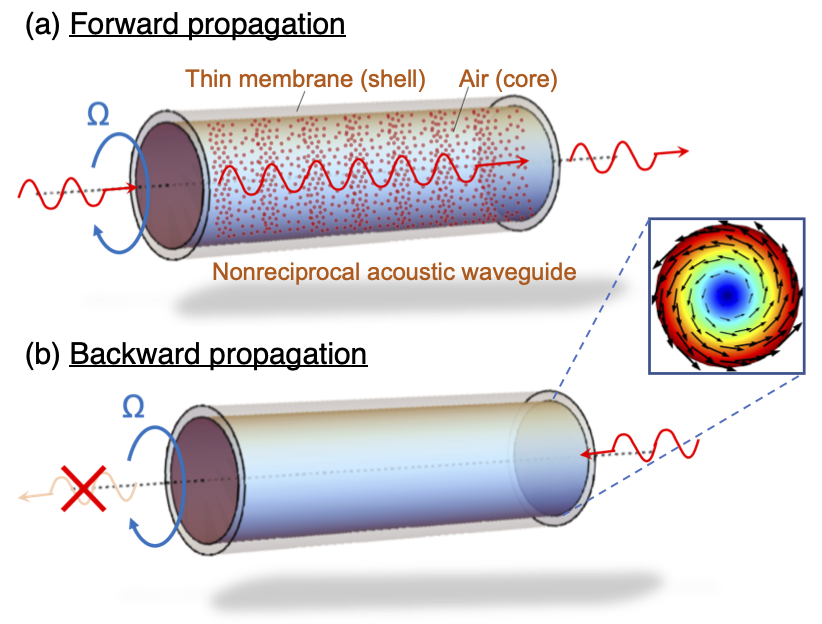}
    \caption{Scheme of a spinning column of air with spinning angular frequency $\Omega$ (in rad/s), where (a) shows the forward allowed propagation while (b) depicts the backward forbidden scenario. A thin-membrane is schematized to separate the spinning and air at rest; the dots schematize the air particles under the background acoustic field. The black arrows of the inset show the COMSOL-computed \cite{comsol} profile of the background flow velocity field [red (blue) color corresponds to high (low) velocity].}
    \label{fig:fig-scheme}
\end{figure}

In this Letter, we propose a new mechanism for acoustic waveguiding, namely the acoustic spinning fiber (ASF) by leveraging the spinning effect of the fluid inside the ASF (See Fig.~\ref{fig:fig-scheme}). The spinning induces an acoustic effective refractive index greater than one ($n_\textrm{eff}>1$) for some multipolar modes, representing an important step beyond the naturally occurring materials. We consider an acoustic fiber of the same physical properties as the surrounding air (an impedance-matched thin-membrane may isolate the fiber from the exterior: This can be realized for example via a sound-permeable thin-sheet solid material \cite{dong2020bioinspired}). We then show in particular that the multipoles $m=+1$ and $m=+2$ can be used for confinement and waveguiding and thus realizing the equivalent of the optical fiber. Analytical and numerical \cite{SM} models (based on the linearized Navier-Stokes interface of COMSOL Multiphysics \cite{comsol}) further demonstrate the robustness of this concept with respect to leaking and may build the basis towards several exciting new functionalities requiring airborne acoustic waveguiding. We further show the spinning-induced unidirectionality of this guiding process with promising applications such as airborne sound-based communication.\\



\begin{figure}[b!]
    \centering
    \includegraphics[width=1\columnwidth]{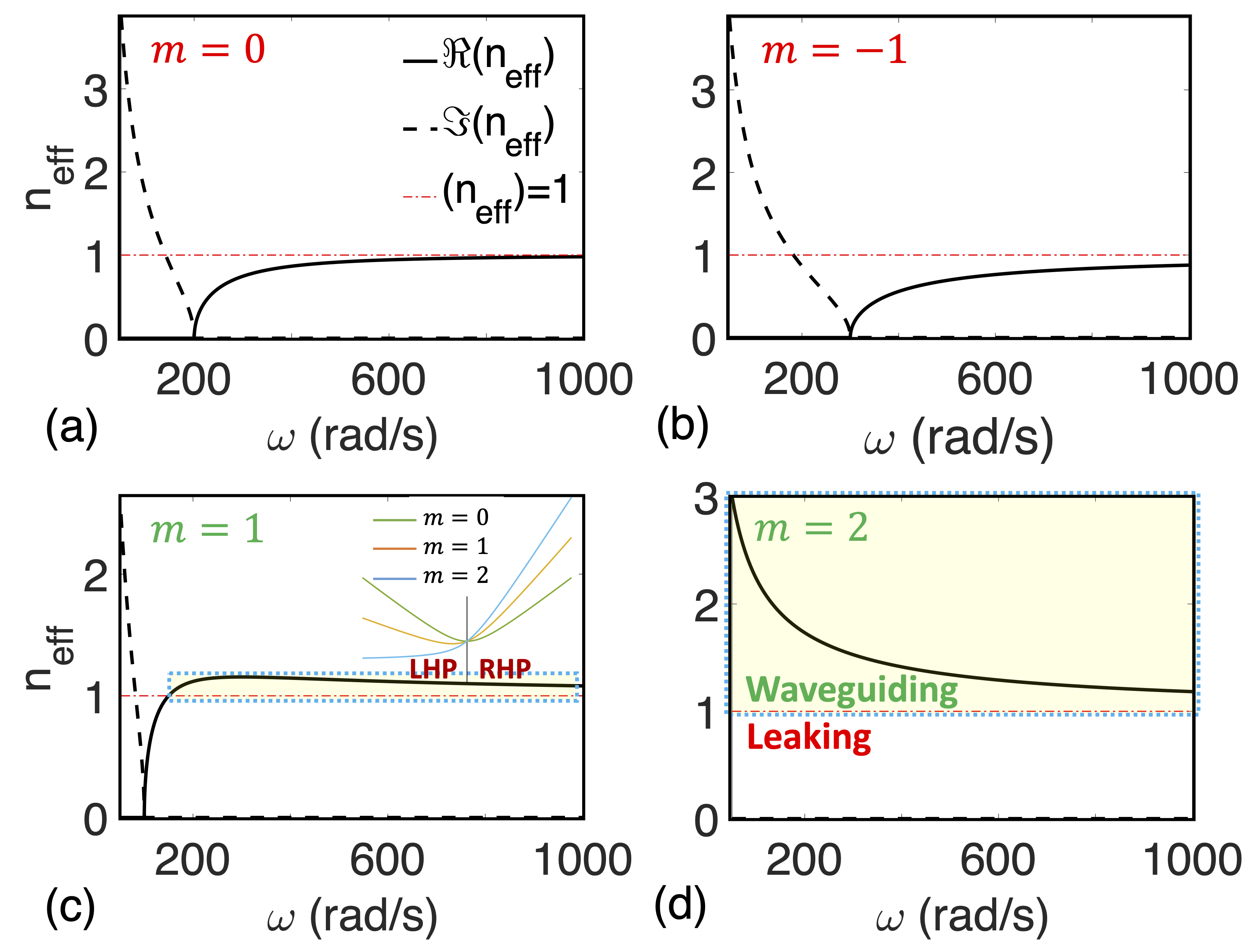}
    \caption{Spinning column of fluid with spinning angular frequency $\Omega=-100$ rad/s. The negative (positive) sign for spinning angular frequency means that the spinning is clockwise (anticlockwise). The few orders dispersion relations are in this case shown for (a) $m=0$, (b) $m=-1$, (c) $m=1$, and (d) $m=2$. The continuous (dashed) lines give the real (imaginary) parts of the acoustic effective index. The upper (lower) region with respect to the red dotted-dashed lines correspond to waveguiding (leaking or radiation). The yellow transparent boxes show the waveguiding regime. The inset in (c) gives the frequency as function of $\Omega/k_0$ for symmetrical ($m=0$) and asymmetrical ($m=1,2$) left hand and right hand propagation (LHP and RHP, respectively).}
    \label{fig:fig-dispmodes}
\end{figure}
\textit{Problem setup.---}Acoustic pressure waves in a spinning nonviscous fluid (air in this study), as schematized in Fig.~\ref{fig:fig-scheme}, obey a Helmholtz-like wave equation (See Supplementary material (SM) for the detailed derivation \cite{SM}) with an equivalent spinning (azimuthal-dependent, i.e., $m$) wavenumber $k_{m}^2=-\left(4\Omega^2+\gamma_{m}^2\right)/c^2$, with $c$ the speed of sound and $p$ the pressure field. $\Omega$ denotes the spinning angular frequency and $\gamma_m=-i(\omega-m\Omega)$ is the azimuthal Doppler angular frequency, with $m\in\mathbb{Z}$. When $\Omega\rightarrow0$, we recover $k_m=\omega/c$, $\forall m$. The features of the spinning wavenumbers $k_m$ were characterized in Ref.~\cite{farhat2020scattering} and further in the SM \cite{SM}. Since $\gamma_{m}$ is a complex-valued parameter, $k_{m}$ has both propagating (real  part) and evanescent (imaginary part) components. There are also frequencies $\omega=(m\pm2)\Omega$ where $k_m\approx0$, reminiscent of near-zero-index metamaterials, as shown in Fig.~\ref{fig:fig-dispmodes}. 
As an illustration, a few multipole mode dispersions in terms of the effective refractive index ($n_\textrm{eff}=k_m/k_0$) as functions of frequency are plotted in Fig.~\ref{fig:fig-dispmodes}. Here, the angular frequency $\Omega=-100$ rad/s (clockwise). It is observed that for low frequencies the imaginary part dominates while the real part is zero. For a given mode's order $m$, the imaginary part vanishes at a specific frequency while the real part increases and converges towards a specific value. In the low frequency regime, there is thus no propagation due to finite imaginary part of $n_\textrm{eff}$. Regarding the propagation regime, we are interested in modes with effective index greater than 1. For $m=0$ or $m=-1$, such condition is not satisfied. However, for $m=1$ and $m=2$ a markedly different behavior is noted. For instance, for $m=1$ starting from a cutoff angular frequency ($\Omega=-100$ rad/s, i.e., a spinning frequency of $|\Omega|/2\pi\approx16$ Hz) we have $n_\textrm{eff}>1$ but the mismatch $n_\textrm{eff}-1\ll 1$. For $m=2$, $n_\textrm{eff}-1>1$ especially for lower frequencies. Hence, for both $m=1$ and $m=2$, confinement of the sound wave and waveguiding are both possible.



Consider a circular air waveguide wrapped by a thin impedance-matched membrane (as an artificial separation of the spinning fluid and the fluid at rest, while being transparent to sound) as schematized in Fig.~\ref{fig:fig-scheme}. On the interface between the spinning waveguide and the host medium, proper continuity conditions need to be satisfied. Namely, the pressure field $p$ and the normal displacement $\zeta_{r}^{m}$ (See SM \cite{SM} for its expression) are continuous. Suppose the radius of the waveguide is $r_1$, inside which spinning air is denoted with subscript 1 and the surrounding air at rest is denoted by subscript 2. We assume that these media are nonviscous and ignore the effects due to friction as the system is airborne. This kind of system models a step-index ASF.

\begin{figure*}[t!]
    \centering
    \includegraphics[width=1.25\columnwidth]{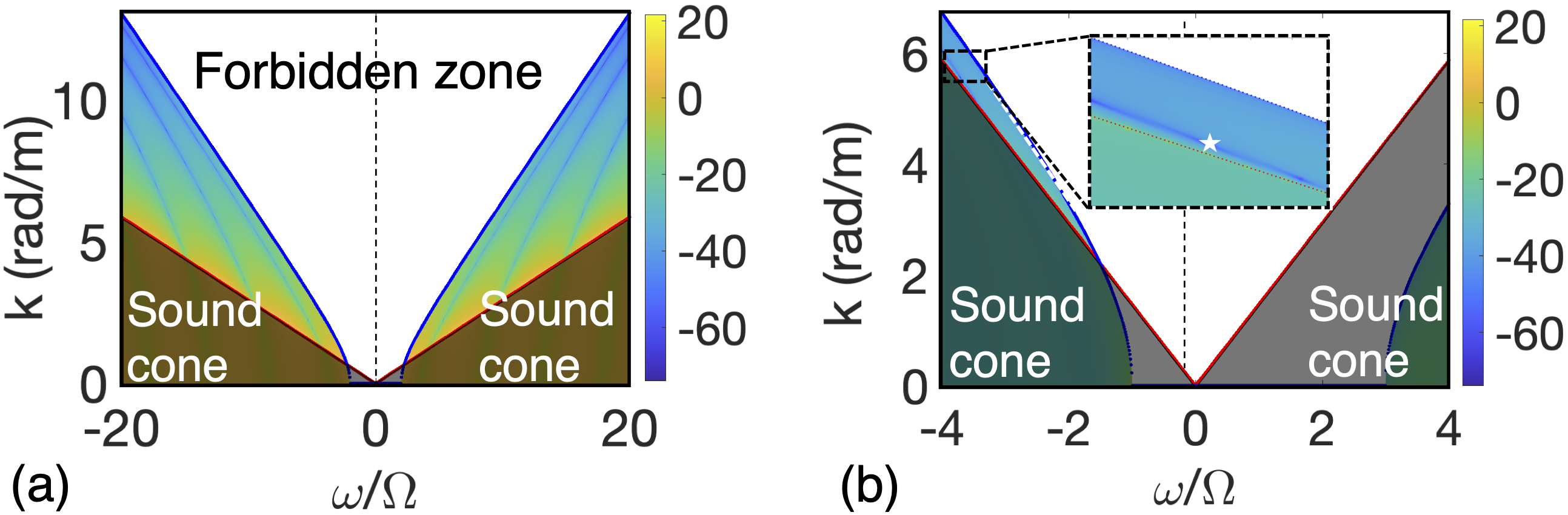}
    \caption{Contourplot of the dispersion relation of the mode (a) $m=0$, i.e., $10\times\log_{10}(|F_0(\omega,k)|)$ and (b) $m=1$, i.e., $10\times\log_{10}(|F_1(\omega,k)|)$ vs $\omega/\Omega$ and the propagation constant $k$, represented by Eq.~(\ref{eq:det_n}).}
    \label{fig:fig-nonrecip}
\end{figure*}
As schematized in Fig.~\ref{fig:fig-scheme}, in each medium ($r<r_1$ and $r\geq r_1$) the pressure field obeys the Helmholtz equation, but with different transverse wavenumbers $\kappa_1$ and $\kappa_2$. The pressure fields have to be finite at $r\rightarrow 0$ indicating $p_1=a_mJ_m(\kappa_1r)e^{im\phi}$ for $r<r_1$, and satisfy the Sommerfield radiation condition, meaning $p_2=b_mH_m^{(1)}(\kappa_2r)e^{im\phi}$ for $r\geq r_1$, for the mode of order $m$. The unknown coefficients $a_m$ and $b_m$ may be deduced from application of the boundary conditions. In passive media (no loss or gain, which is assumed in this study), $\kappa_1^2$ and $\kappa_2^2$ are purely real numbers. To confine propagation inside the waveguide (or ASF, i.e., for $r<r_1$) we have to enforce $\kappa_1=\kappa>0$ and let $\kappa_2=i\beta$ ($\beta$ is a positive real number) such that the pressure field decays far-off from the ASF. $\kappa$ and $\beta$  are given by
\begin{equation}
    \kappa^2 = \frac{\omega^2}{c_1^2}\left[\left(m\alpha-1\right)^2-4\alpha^2\right]-k^2\, ,\, \;\; \beta^2 = k^2-\frac{\omega^2}{c_2^2}\, ,
  \label{eq:trans_k}
\end{equation}
where $c_1$ and $c_2$ are the acoustic wave speed in the ASF and in the surrounding air, respectively; $\alpha=\Omega/\omega$ is the dimensionless spinning ratio and $k$ is the propagation constant of the waveguided mode inside the ASF. The peculiar dispersion inside the spinning medium arises from the unique governing equation. The fields can thus be re-written using the modified Hankel functions, i.e., $p=a_mJ_m(\kappa r)e^{im\phi}$ for $r<r_1$ but $p=b_mK_m(\beta r)e^{im\phi}$ for $r\geq r_1$ (where we choose to keep the notation of $b_m$ although it is proportional by a factor of $(\pi/2)i^{m+1}$). This choice ensures that the acoustic wave does not propagate energy (leaking) towards the exterior of the ASF, as the fields decay exponentially with $r$ ($\lim_{r \to +\infty}|p|=0$).\\




\textit{Spinning-induced acoustic fiber.---}The $m=0$ mode does not result in $n_\textrm{eff}>1$ and therefore is not a propagating mode, as seen in Fig.~\ref{fig:fig-dispmodes}(a). Yet, it is interesting to briefly analyze this fundamental mode to understand the basic effects. We suppose that a column of a fluid with $n_1=\sqrt{5}$ and of radius $r_1$ is spinning and we apply the convenient boundary conditions at $r=r_1$. Although for this fundamental mode waveguiding is not induced by spinning but rather by $n_1>1$ inside the fiber, spinning still has some intriguing effect on the dispersion [as seen in Fig.~\ref{fig:fig-nonrecip}(a)] and the induced waveguiding overall (See Section III and Fig.~1 in SM \cite{SM} for a detailed discussion of the fundamental mode-based waveguiding mechanism). For the higher order mode, i.e., we investigate acoustic waveguiding solely induced by spinning and thus not requiring a refractive index greater than one. We consider a medium of the same properties as the surrounding air and we analyze the waveguiding effect. Hence, it is obvious that the waveguiding mechanism differs drastically from the previous ones (including optical waveguiding). Here, the confinement of the acoustic fields inside the ASF is not induced by the increased refractive index but rather by the spinning dynamics. This is manifested by Eq.~(\ref{eq:trans_k}) that can lead to
\begin{equation}
\kappa^2+\beta^2=\frac{\omega^2}{c^2}\left[\left(m\alpha-1\right)^2-4\alpha^2-1\right]\, ,
\label{eq:disp_nb}    
\end{equation}
if $c_1=c_2=c_\textrm{air}=c$.

To allow guided modes, both $\kappa$ and $\beta$ should be real numbers (so to ensure propagation in the core for $\kappa_1=\kappa$ and exponential decay in the surrounding via $\kappa_2=i\beta$) which requires that the RHS of Eq.~(\ref{eq:disp_nb}) should be positive. When we allow different material properties ($n_1>n_2$, i.e., $c_1<c_2$) without spinning, this is just $1/c_1^2-1/c_2^2>0$. But in Eq.~(\ref{eq:disp_nb}) this condition becomes $(m\alpha-1)^2-4\alpha^2\geq1$ or equivalently $\alpha\left[\left(m^2-4\right)\alpha-2m\right]\geq0$. This inequality explains the findings of Fig.~\ref{fig:fig-dispmodes} where waveguiding was only possible for $m=1$ or $m=2$, for $\Omega=-100$ rad/s (See Figure 3 of SM for details \cite{SM}).
Now, by applying the more general boundary condition for the multipole $m$, we can get the dispersion equation 
\begin{equation}
\begin{vmatrix}
J_m(\kappa r_1) & -K_m(\beta r_1)\\
\frac{\left(2\Omega^2-\gamma_{m}^2\right)\kappa J_m'\left(\kappa r_1\right)-\frac{3\gamma_{m}\Omega im}{r_1} J_m\left(\kappa r_1\right)}{\rho_1\left(4\Omega^2+\gamma_{m}^2\right)\left(\Omega^2+\gamma_{m}^2\right)} & -\frac{1}{\rho_2\omega^2}\beta K_m'(\beta r_1) 
\end{vmatrix}=0\, ,
\label{eq:det_n}    
\end{equation}
with both $\kappa$ and $\beta$ satisfying Eq.~(\ref{eq:disp_nb}), and by denoting $F_m(\omega,k)$ the LHS of Eq.~(\ref{eq:det_n}).

The dispersion resulted from this model is depicted in Fig.~\ref{fig:fig-nonrecip}(b) for the case $m=1$ (See Fig.~\ref{fig:fig-dispmodes}(c) for the medium's dispersion), exhibiting unique behaviors that were not previously observed in common waveguiding systems. In fact, the red solid line indicates the limit of the sound cone, i.e., $\omega/c$. The modes appearing below this line correspond to leaking (radiative regime), i.e., acoustic waves that propagate into the surrounding space. The second limit plotted in blue line corresponds to the forbidden zone (neither propagating nor leaking modes are possible above the blue curve, i.e., the white region of the 2D plot). This curve has the specific variation $(\omega/c)\sqrt{(\alpha-1)^2-4\alpha^2}$. Unlike classical waveguiding schemes, this forbidden zone is nonlinear. It does not start from zero but rather has cutoff frequencies due to the imaginary part of $k_n$ (i.e., solutions of $(\alpha-1)^2-4\alpha^2=0$). This cutoff frequency depends on the modes, indicating  tunability of acoustic waveguiding  (negative and positive spinning lead to the same effect). Hence, it is confirmed that a simply spinning column of air results in confinement and waveguiding [dark blue region of the 2D plot, see inset of Fig.~\ref{fig:fig-nonrecip}(b)]. Moreover, it is worth noting that such confinement is asymmetric, i.e., it results in the one-way waveguiding effect. For instance, Fig.~\ref{fig:fig-nonrecip}(b) shows the different dispersion with opposite sign of the spinning direction. This shows that sound propagating towards the right and the left are markedly distinct. For some frequencies highlighted in Fig.~\ref{fig:fig-nonrecip}(b), it is possible to enforce unidirectional propagation, where RHP is possible while LHPis prohibited. Such feature is not possible with classical waveguides and is reminiscent of spin-induced waveguiding \cite{slussarenko2016guiding} or vortex beams \cite{zou2020orbital}.\\ 

\textit{Mechanism of the ASF.---}To explain the intriguing findings discussed above (nonreciprocal feature), we define state vectors as $\ket{\psi}=(p,{\bf v})^T$, with $(\cdot)^T$ the transpose operator and ${\bf v}=(v_r,v_\phi)^T$ (the $z$-component is decoupled), endowed with scalar product $\braket{\psi_1,\psi_2}=\int_\mathcal{S} dS\left\{p_1^*p_2+{\bf v}_1^*\cdot{\bf v}_2\right\}$, with $\mathcal{S}$ the transverse surface of the ASF. In this way, the governing equation (5) of the SM \cite{SM} can be simply written in the more elegant way $\left[H^{(0)}+\delta H\right]\ket{\Psi}=\omega\ket{\Psi}$, with $\omega$ the eigenfrequency of the system. $H^{(0)}$ ($\delta H$) is the time-evolution operator of the system in the absence of bias (perturbation operator) (see SM for the expressions \cite{SM}). It can be shown that in the absence of bias ($\Omega=0$) two modes exist ($m=\pm 1$) that have the same eigenfrequency $\omega$, i.e.,
\begin{equation}
\ket{\pm}= \alpha_\pm\zeta_\pm \begin{pmatrix}
i\rho\omega J_\pm1\left(\kappa r\right)\\
\kappa_1 J'_\pm1\left(\kappa r\right)\\
\frac{\pm i}{r}J_\pm1\left(\kappa r\right)
\end{pmatrix} \, ,
\label{eq:eq-eigenp3}    
\end{equation}
with the constants $\zeta_\pm$ the renormalization factors, computed by using the defined scalar product and where $\alpha_\pm=e^{i(kz\pm\phi)}/i\omega\rho$.

When we turn spinning on (biased system, i.e., $\Omega\neq0$), we make the assumption that the new eigenvectors lie in the subspace described by the unbiased waveguide modes given in Eq.~(\ref{eq:eq-eigenp3}). We can thus write the new eigenvectors as a linear combination of $\ket{\pm}$, with some unknown complex coefficients $\mu^+$ and $\mu^-$, i.e., $\ket{\psi}=\mu^+\ket{+}+\mu^-\ket{-}$. By using the fact that $H^{(0)}\ket{\pm}=\omega_0\ket{\pm}$, we obtain
\begin{equation}
\begin{pmatrix}
\bra{+}\delta H\ket{+} & \bra{+}\delta H\ket{-}\\
\bra{-}\delta H\ket{+} & \bra{-}\delta H\ket{-}
\end{pmatrix}
\begin{pmatrix}
\mu^+\\
\mu^-
\end{pmatrix}
=\left(\omega-\omega_0\right) \begin{pmatrix}
\mu^+\\
\mu^-
\end{pmatrix}\, ,
\label{eq:eq-pert2}    
\end{equation}
where we have $\bra{+}\delta H\ket{-}=\bra{-}\delta H\ket{+}=0$ and that $\bra{+}\delta H\ket{+}=-\bra{-}\delta H\ket{-}$ \cite{SM}. Hence, the eigenvalues of the biased system (spinning on) are given by virtue of Eqs.~(\ref{eq:eq-eigenp3})-(\ref{eq:eq-pert2}) as
\begin{equation}
\omega^{\pm}=\omega_0\pm\bra{+}\delta H\ket{+}\, ,
\label{eq:eq-pert3}    
\end{equation}
where $\bra{+}\delta H\ket{+}=\Omega$ \cite{SM} for the modes that we consider in this demonstration, i.e., $m=\pm 1$. This analysis demonstrates that the spinning of air breaks the reciprocity similarly to the application of a magnetic field in optics (Zeeman effect) \cite{kupriyanov2004antilocalization}, and hence lifts the degeneracy of eigenfrequencies, corresponding to both directions of propagation, shown by Eq.~(\ref{eq:eq-pert3}). But this can occur only if $m\neq0$, as for $m=0$ the dispersion is symmetrical, which is verified in Fig.~\ref{fig:fig-nonrecip}(a) and predicted by the dispersion shown in the inset of Fig.~\ref{fig:fig-dispmodes}(c).

\begin{figure}[t!]
    \centering
    \includegraphics[width=1\columnwidth]{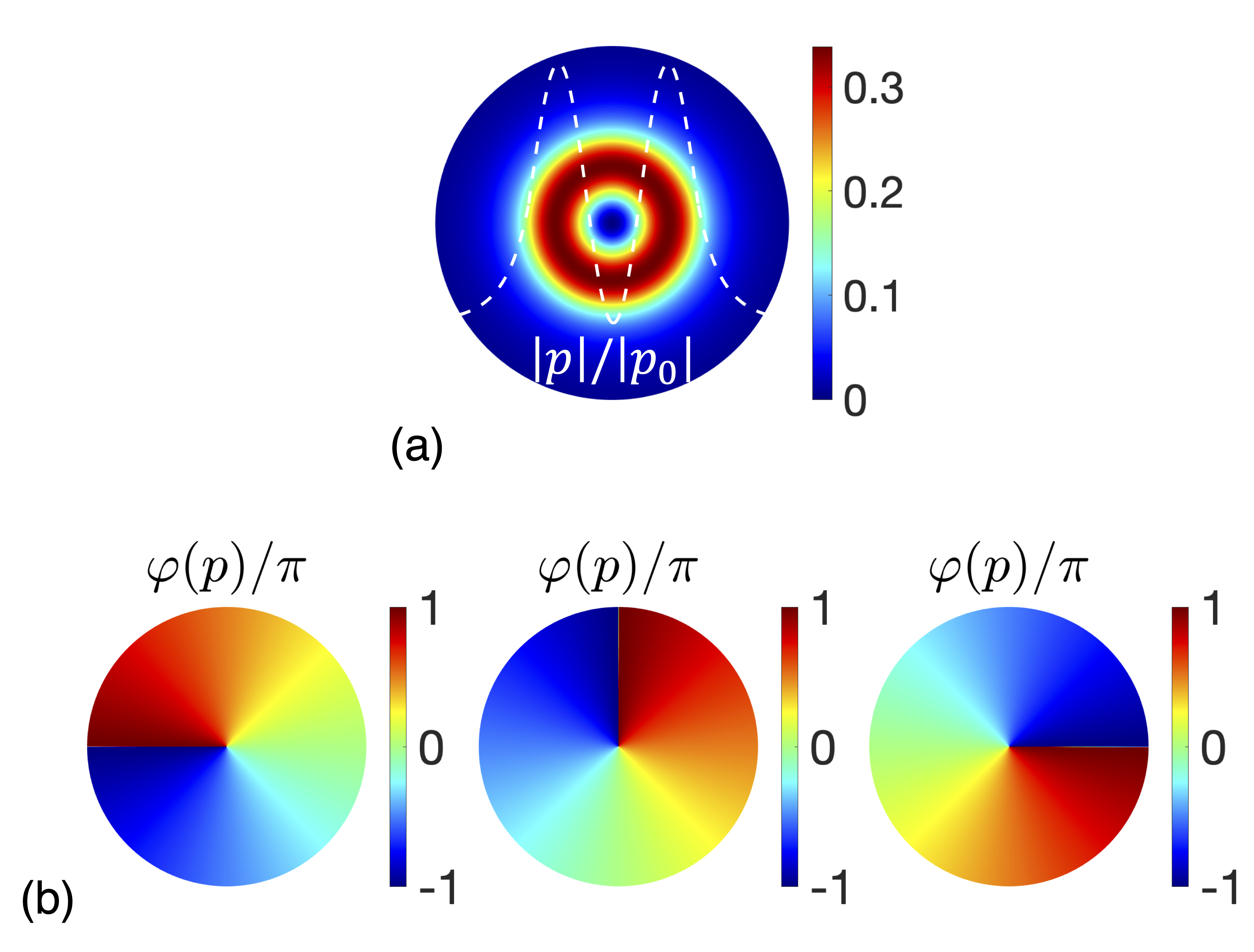}
    \caption{(a) Snapshot of the normalized amplitude of the pressure field $|p|/|p_0|$ inside the ASF and in the outside at the frequency highlighted by a white star in (a). The white dashed curve plotted on top of it is the cut-line along the direction $\phi=0$. (b) Snapshot of the phase [$\varphi(p)$ in units of $\pi$] of the pressure field $p$ inside the ASF and in the outside at three different locations in the propagation direction ($z$-axis), i.e., from left to right: $z=0,\pi/2k$, and $\pi/k$.}
    \label{fig:fig-nonrecip2}
\end{figure}

In the same vein, Fig.~\ref{fig:fig-nonrecip2}(a) depicts the pressure field distribution, which not only shows the confinement, but also exhibit a minimum at the center of the ASF, indicating the presence of a singularity. Similar feature was observed previously for vortex beams, carrying an acoustic orbital angular momentum (OAM) \cite{zou2020orbital}. In our case, as was shown recently in Ref.~\cite{farhat2021transverse}, spinning fluids carry also a spin and orbital angular momentum features. This may explain the singularity at the center of the ASF seen in Fig.~\ref{fig:fig-nonrecip2}(a). Moreover Fig.~\ref{fig:fig-nonrecip2}(b) gives the phase of the pressure field (in units of $\pi$) at the point highlighted by a white star in Fig.~\ref{fig:fig-nonrecip}(b) for different locations along the propagation axis ($z$-axis here), i.e., for $z=0,\pi/2k$, and $\pi/k$, where $k$ is the propagation constant taken from Fig.~\ref{fig:fig-nonrecip}(b). This further confirms the orbital nature of this waveguiding phenomena.\\

\textit{Discussion and conclusion.---}In this Letter, we show for the first time spinning-induced acoustic waveguiding. We leverage the rotation of a spinning column of air (surrounded by air, too) to obtain intriguing azimuthal-dependent dispersion. For specific parameters, we analyze the condition of confinement of airborne sound and describe in details the resulted acoustic spinning fiber. In particular, the suggested mechanism is endowed with several intriguing properties, such as (i) unidirectionality of the propagation due to the asymmetry in the dispersion of the waveguide modes and the acoustic analogue of Zeeman effect, (ii) presence of a singularity of the pressure field reminiscent of acoustic OAM, and (iii) tunability of the waveguiding by tuning or flipping the sign of the spinning angular frequency. Moreover, the proposed ASF is the acoustic counterpart of the optical fiber, for which one of the main advantages is the enhanced bandwidth and low loss compared to hollow electric waveguides \cite{singal2016optical}. Thus, the proposed ASF may possess an enhanced bandwidth compared to duct waveguides and offer longer propagation distances, as it does not suffer from absorption loss or undesired reflections (reflection channel is forbidden by nonreciprocal behavior). All these properties demonstrate the potential of this ASF that can be applied in several domains including airborne sound communication (by avoiding the undesired parasitic back-reflection) or to shed new light onto the physics of acoustic vortex beams and nonreciprocal physics.

Although the experimental demonstration of this concept is outside the scope of the present Letter, some recent experimental studies indicate that our proposal may be demonstrated (See SM \cite{SM}) using similar configurations \cite{fleury2014sound,quan2021odd}. Moreover, the COMSOL-based \cite{comsol} simulations shown in the SM further demonstrate the accuracy of our model \cite{SM}\\


\textit{Acknowledgements.---}This work was supported by King Abdullah University of Science and Technology (KAUST) Office of Sponsored Research (OSR) under Grant No. OSR-2020-CRG9-4374, as well as KAUST Baseline Research Fund BAS/1/1626-01-01.


\nocite{*}

\providecommand{\noopsort}[1]{}\providecommand{\singleletter}[1]{#1}%

\end{document}